# Anisotropy in vortex phase diagram and the pinning force density in the basal plane of $YNi_2B_2C$

Pradip Das[a], C.V. Tomy[a], H. Takeya[b], S. Ramakrishnan[c] and A.K. Grover[c]

[a]Department of Physics, Indian Institute of Technology Bombay, Mumbai-400076, India
[b]National Institute of Materials Science, Ibaraki 305-0047, Japan
[c]Department of Condensed Matter Physics & Materials Science, Tata Institute of Fundamental Research, Mumbai 400 005, India

## ABSTRACT

We present magnetic field dependence of the critical current density from dc-magnetization measurements concerning the anisotropic behavior of flux line lattice (FLL) in a single crystal of $YNi_2B_2C$. The peak effect (PE) phenomenon is observed for all crystallographic orientations, but the second magnetization peak (SMP) anomaly is observed only for $H \parallel a$. Our study reveals that the FLL obtained when $H \parallel [110]$ is better ordered within the basal plane. However, the FLL for $H \parallel c$ is found to be even more ordered than that for $H \parallel [110]$. The perfect square symmetry of the FLL for $H \parallel c$ is perhaps responsible for promoting the realization of the best spatial order of the FLL prior to the onset of PE, indicating a correlation between the crystalline lattice and the FLL. We have also found a change over in the power law governing the decay of the critical current density which is identified as a crossover from weak to weaker pinning regime in the phase diagram.

*Key words:* Flux line lattice, Pinning force, Peak effect, Anisotropy

# 1. INTRODUCTION

The study of the interplay between the symmetry of crystal lattice (CL) and flux line lattice (FLL) has assumed accentuated importance in recent years in the context of vortex phase diagram studies in superconductors. Angular dependent magnetization hysteresis studies in single crystals of $V_3Si$ [1] having different morphology demonstrated that the quality of the ordered Bragg glass (BG) phase could crucially depend on the orientation of the magnetic field with respect to the underlying crystallographic lattice (CL), even though the crystalline symmetry of $V_3Si$ is cubic and the anisotropy in its intrinsic superconducting parameters is marginal. In $V_3Si$ the best ordered BG phase, with least pinning, gets formed when the magnetic field is oriented along [110] and the single crystal is spherical in shape, with uniform demagnetization factor. A correlation between the angular dependence in spatial order-disorder transition at high fields and that in the change in local symmetry of the FLL at low field was also supported by the observation of the transition to from the hexagonal to the square FLL in a $V_3Si$ crystal for $H \parallel [110]$ by Yethiraj $et\ al$ [2] from small angle neutron scattering (SANS) measurements. The notion of correlation between the symmetry of CL and that of the FLL has also been evoked by Park $et\ al.$ [3], while analyzing and reporting the angular variation in field dependence of heat capacity data in single crystals of a borocarbide superconductor, $LuNi_2B_2C$. Park $et\ al.$ [3] found an anomalous behaviour in the field dependence of the heat capacity measurements in the basal plane of the tetragonal structure of $LuNi_2B_2C$. The threshold field for the said anomaly for $H \parallel [110]$ was observed to be lower than that for $H \parallel [100]$, and it was surmised [3] that this difference reflected the delay in realization of the square FLL for $H \parallel [100]$. While carrying out the angular dependence study of the peak effect (PE) phenomenon in a single crystal of $LuNi_2B_2C$, Jaiswal-Nagar $el\ al.$ [4] have further reported that the threshold field for the onset of PE ($H_p^{on}$) is larger for $H \parallel [110]$ as compared to that for $H \parallel [100]$, thereby implying that the quality of spatial order in the BG state before the onset of the PE for $H \parallel [110]$ is superior to that for $H \parallel [100]$. On the basis of another study on a single crystal of $YNi_2B_2C$ (YNBC), Jaiswal-Nagar $et\ al.$ [5] had stated that the state of spatial order in the BG state for $H \parallel c$ is superior to that for $H \perp c$. The distinction between the different orientations in the basal plane was not brought out in the data of Ref. 5. Though a number of experimental studies exist on the correlation between the FLL and the CL, there is still a justification for motivated studies to seek specific answers in a chosen system. In this spirit, we have explored the angular dependence of the anomalous variation in magnetization hysteresis width for field oriented along different directions in the basal plane of a single crystal of YNi$_2$B$_2$C, which is parallelepiped in shape. The new results being reported here fortify the assertions that (i) the BG state is most weakly pinned, when the vortex lattice conforms to a perfect square symmetry (i.e., for $H \parallel c$), (ii) the correlation volume for the ordered state prior to the PE is larger for $H \parallel [110]$ as compared to that for $H \parallel [100]$ and (iii) when the vortex lattice is quasi-square (as for $H \parallel [100]$), smaller demagnetization factor promotes the realization of better spatial order prior for vortex state to the PE.

## 2. EXPERIMENTAL DETAILS

The YNi$_2$B$_2$C crystal used for the present study is in the shape of a parallelepiped with dimensions 3.5x0.7x0.67 mm$^3$, and is a portion of the crystal used for the point contact Andreev reflection studies [6] earlier. The crystal has a $T_c$ of 15.5 K and its longest dimension (3.5 mm) is cut along the [100] direction, as shown in the inset of Fig. 1. The other two smaller dimensions, 0.7 mm and 0.67 mm, of the parallelepiped are aligned in [010] and [001] direction, respectively. The dc magnetization hysteresis measurements were performed using a vibrating sample magnetometer (VSM) option in a Physical Property Measurement System (PPMS, Quantum Design Inc., USA). The magnetic field was ramped at a rate of 100 Oe/s in all the magnetization measurements. To study anisotropy in the basal plane, the crystal was rotated with respect to the magnetic field within the basal plane from [100] direction, designated as the *long* direction, to [010] direction, designated as the *short* direction, through [110] direction (body diagonal of the basal plane) in increments of 15° ($\Delta\theta$). Needless to add, for the tetragonal structure, the [100] and the [010] directions are crystallographically equivalent. To study the out-of-basal plane anisotropy, the crystal was rotated with respect to the magnetic field from [100] (*long*) direction to [001] direction (*c*-axis). We recorded magnetic hysteresis loops at several temperatures in the superconducting state in each crystal orientation.

# 3. RESULTS AND DISCUSSION

## A. Angular dependence in the critical current density

Figure 1 displays a typical magnetization hysteresis loop (only the second and the fifth quadrants are shown for clarity) recorded at 2 K with $H \parallel [100]$ (long direction). In order to visualize the two distinct features clearly, namely, the peak effect (PE) and the second magnetization peak (SMP), we show the selected expanded portions of the magnetization loop in the insets (a) and (b) of Fig. 1, respectively. The PE can clearly be seen in both the forward leg and the reverse leg of the loop where as the SMP is resolved only in the reverse leg of the loop (cf. inset (b), Fig. 1). *Prima-facie*, the hysteresis loop is asymmetric in nature which may be caused by the thermomagnetic history dependence of the critical current density $j_c$ (proportional to half width of the hysteresis, $M(H) - M_{FC}(H)$ [7,8,9], where $M_{FC}(H)$ is the field cooled magnetization at a given $H$), arising from the differences in the residual disorder in FLL and/or the difference in surface pinning in the forward and reverse legs of the magnetization loop [10].

Figures 2(a) and 2(b) show the portions of $M$-$H$ isotherms recorded at two typical temperatures, viz., 2 K and 6 K. These figures also include the behaviour of $M$-$H$ isotherms, when the crystal is rotated with respected to the applied field from [100] (i.e., from long *a*-direction) to [010] (i.e., to the short *a*-direction) through the basal plane [110]. The PE anomaly in the magnetization loops is clearly visible for all the orientations (see the insets, where the plots across the PE region are shown on an expanded scale). The peak position of the PE ($H_p$) and the width of the PE region depend on the orientation of the crystal with respect to the magnetic field; both decrease as one goes from the [100] direction (long *a*) to the [110] direction, and, thereafter, increase further as the crystal gets rotated back to the (short) *a*-direction, i.e., [010] direction. Since the PE is associated with an order-disorder transition [11], a sharper PE or a smaller width of the PE region implies a narrower region over which the order-disorder transition occurs. Hence, one can surmise that the well ordered FLL for [110] direction occupies a larger $H$ region in the vortex phase diagram as compared to that for the [100] or [010] directions. The second magnetization peak (SMP) can clearly be resolved only for $H \parallel [100]$

(e.g., see inset (b) of Fig. 1) and $H \parallel [010]$ (Fig. 2) directions. It is believed that the SMP anomaly, occurring much prior to the PE, is primarily due to the induced pinning [12]. The absence of SMP anomaly in [110] direction as compared to its presence in [100] or [010] directions in $YNi_2B_2C$ is consistent with a similar observation in $LuNi_2B_2C$ [4] and again supports the view that the FLL for $H \parallel [110]$ is better spatially ordered as compared to that for $H \parallel [100]$ or for $H \parallel [010]$.

To further corroborate the possibility of correlation between the FLL and the CL, we plot in Fig. 3, the normalized critical current density $j_c$ derived from the magnetic isotherms at 2 K, as a function of reduced field ($h = H/H_{c2}$), for different crystal orientations within the basal plane. We have normalized $j_c$ with respect to two positions in the vortex phase diagram of type-II superconductors; w.r.t. the peak position of the peak effect ($H_p$) where the disorder in FLL is maximum and w.r.t. the low field region where the FLL is in an ordered state. In Fig. 3(a), we show $j_c$ normalized with respect to the $j_c$ at $H_p$ [1]. In the low field regime ($h < 0.02$), $j_c$ is weakly dependent on the field for all the orientations thereby implying that the field induced interaction (i.e., collective pinning [13,14]) effect is small and the vortices are probably individually pinned. The low field regime is often designated as the small bundle pinning (SBP) region [15]. As the applied field increases, $j_c$ starts to depend on the field and appears to display a power law, i.e., $j_c \propto h^{-n}$, in all the crystal orientations within the basal plane. We demarcate the entire field range prior to the onset of the PE into two different power law regions in Fig. 3(a) (for brevity we have identified the intervals only for the $H \parallel [100]$. Two different power laws governing the decay of the critical current may be indicative of two different pinning regimes of the FLL. A comprehensive explanation regarding this phenomenon shall be given in a subsequent subsection. Here, it is useful to calculate the relative change in the correlation volume $V_c$ ($j_c \propto 1/\sqrt{V_c}$) within the Larkin-Ovchinnikov (L-O) theory of collective pinning [13,14] from the $j_c(h)$ data of Fig. 3(a). The ratio of $j_c$ at $H_p$ to that at the onset of the PE ($H_p^{on}$) is about 1/5 for $H \parallel [100]$, while the same ratio is about 1/50 for $H \parallel [110]$. This implies that the correlation volume for $H \parallel [100]$ orienta-tion shrinks only to about 1/25 of its value at the onset of PE whereas for $H \parallel [110]$, $V_c$ at $H_p$ has shrunk to 1/2500 of its value at $H_p^{on}$. This in turn once again implies a better spatially ordered FLL for $H \parallel [110]$ as compared to that for $H \parallel [100]$. $j_c$ in the region above the peak position of the PE has been known to show no directional dependence, as reported in an another low $T_c$ superconductor V$_3$Si [1]. In Fig. 3(a), if we further compare the ratio of $j_c$ at peak position of the PE to that at its onset position for $H \parallel [100]$ and $H \parallel [010]$, we note that for the latter orientation, the ratio is smaller by a factor of 1.8. This implies that for long $a$-direction, a better spatial order is promoted prior to the onset of PE due to its smaller demagnetization factor. Normalizing $j_c$ w.r.t. its value at the low field limit (i.e., the SBP region) also essentially brings about a similar behaviour (Figure 3 (b)). While noting that $j_c$ decreases as the field increases, we can ascertain that at the onset of PE field, $j_c$ takes its lowest value for $H \parallel [110]$ as compared to other orientations. This in turns implies that the correlation volume ($V_c$) is maximum prior to the onset of the PE for $H \parallel [110]$. This rationalize the larger reduction in correlation volume across the PE for $H \parallel [110]$.

We now focus our attention onto the $j_c$ variation for the out of basal plane rotation of the magnetic field, i.e., when the applied field is rotated from $a$-axis ([100]) towards $c$-axis ([001]). In Fig. 4, we show the variation of $j_c$ (normalized with respect to its value at $H_p$) as a function of $h$ at 5 K for three different orientations of the crystal, i.e., $\theta = 0°$ ([100]), 15° and 90° ([001]), where $\theta$ is the orientation between the applied field and the crystal axis. For the sake of comparison, we have included in Fig. 4 the normalized $j_c$ at 5 K for $H \parallel [110]$ (basal plane). The SMP anomaly could easily be resolved only for $\theta = 0°$ ([100]) and 15°. The absence of SMP for $H \parallel c$ implies that the FLL prepared for $H \parallel c$ is weakly pinned as in the case of $H \parallel [110]$. It can also be noted from Fig.4 that for $H \parallel c$, $j_c$ drops steeper in the second power law regime ($h > 0.2$) compared to the other orientations. To compare the quality of weakly pinned FLL along $H \parallel [001]$ and $H \parallel [110]$, we have calculated the relative change in the correlation volume in the interval between $H_p^{on}$ and $H_p$ w.r.t. the same for $H \parallel [100]$ orientation. The values of the ratio, $[(V_c^{H_p})/(V_c^{H_p^{on}})]_{001}/[(V_c^{H_p})/(V_c^{H_p^{on}})]_{100}$ and $[(V_c^{H_p})/(V_c^{H_p^{on}})]_{110}/[(V_c^{H_p})/(V_c^{H_p^{on}})]_{100}$ calculated from Fig.4 are 62500 and 36, respectively. These observations along with our previous observations made in the basal plane suggest that the best ordered FLL is formed for $H \parallel [001]$ even though within the basal plane one can find better ordered FLL for $H \parallel [110]$. Here, it is pertinent to mention that in SANS measurements, a perfect square lattice (with apex angle $\beta = 90°$) was observed only for $H \parallel [001]$, whereas for $H \parallel [110]$, the FLL symmetry was only quasi-square (with apex angle $\beta < 90°$) [16,17,18]. This implies that the perfect square lattice promotes the creation of best spatial order prior to the PE.

We have also estimated the upper critical field values from the magnetization hysteresis using the deviation from linearity criterion and found the average out-of plane anisotropy, i.e., $0.5(H_{c2}^{[100]} + H_{c2}^{[110]})/H_{c2}^{[001]}$, to be 1.15, which matches well with the value reported by Metlushko *et al.* [19]

### B. Evolution of pinning force density and the pinning crossover

To investigate the scenario as to whether the pinning force ($F_p$) has a dependence on the crystal directions and temperatures, we have calculated the pinning force $F_p (\propto j_c H)$ for different orientations of the crystal at different temperatures. In Fig.5(a), we show $F_p$ plotted against the reduced field $h$ ($= H/H_{c2}$) for $H \parallel [100]$. In addition to the two expected maxima [16, 20], one in the low field region (due to the usual maximum in the pinning force density) and the second in the high field region (near to normal-superconductor boundary) due to the PE, we observe a third maximum in the intermediate field [16] range ($h \sim 0.2$). The possibility of this intermediate peak occurring due to the SMP can be ruled out since the field values at which the SMP occurs is almost half ($h \sim 0.1$) the present values ($h \sim 0.2$). In order to compare $F_p$ values for different temperatures/orientations, $F_p$ for each temperature/orientation was normalized with respect to its maximum value in the intermediate field range. The irregular discontinuous varia-tion of $F_p$ at low temperatures is associated with the flux jumps in the low field region due to FLL symmetry transitions [21]. We also observe a weak but significant temperature dependence of the second maxima at $h \sim 0.2$ in our $F_p$ vs. $h$ plots (cf. inset panel, Fig. 5(a)). The pinning force behaviour is almost the same for $H \parallel [110]$, but the prominence of the temperature variation is somewhat diminished compared to $H \parallel [100]$ (Fig. 5(b)). Prima-facie, it may be related to the fact that the FLL along the [110] direction is better ordered than the [100] direction as explained in the earlier sections. Figure 5(c) shows normalized $F_p$ as a function of $h$ for different orientations of the crystal at 2 K. There is a clear bump at $h \sim 0.2$ for different orientations of the crystal.

It is relevant here to point out that at $h \sim 0.2$, we have already observed a sharp drop in the critical current and a change in the power law (Fig. 3(a)). From transport measurements, Eskildsen *et al.* [16] have observed a similar hump type maximum in pinning force density at approximately $h \sim 0.2$ with $H \parallel c$ at 2.2 K in LuNi$_2$B$_2$C. They also observed a maximum in $\xi_L / a_0$ vs. field data around $h \sim 0.2$, calculated from SANS measurements, where, $\xi_L$ is the longitudinal correlation length extracted from the width of rocking curve and $a_0$ is the FLL lattice spacing. The decrease of $\xi_L / a_0$ for $h > 0.2$ implies a marked deviation from the Larkin-Ovchinnikov collective pinning model which predicts only an increase for $\xi_L / a_0$ as a function of field. The observed hump at $h \sim 0.2$ is attributed to the anomalous field dependence of the elastic modulii of the square FLL, which gets stabilized by nonlocal correction to the London model. It has been argued that for $h < 0.2$, tilt modulus ($c_{44}$) controls the pining force behavior of the FLL and for $h > 0.2$, the same is governed by shear modulus ($c_{66}$). In another study on a platelet shaped single crystal of YNi$_2$B$_2$C [20], a similar maximum in $F_p$ was observed for $h \sim 0.2$, which was also attributed to the crossover from tilt to shear modulus, along the same assertions made in Ref. [16]. Here, it may be pertinent to state that in Ref. 20, the maximum at intermediate field ($h \sim 0.2$) for $H \parallel c$ was independent of temperature. This temperature independence was justified to the fact that the functional form for calculation of elastic moduli for the square lattice in Ref. 16 included only reduced field parameters with no explicit temperature dependence. The temperature dependence in the pinning force data in the current investigation therefore raises additional/new rationalization. As regards the crossover in pinning behaviour at $h \sim 0.2$, we recall here another assertion made while reporting newer finding in the vortex phase diagram in 2$H$-NbSe$_2$ [22]. It is claimed [22] that in 2$H$-NbSe$_2$, current density displays path dependence prominently at lower $h$, which disappear at intermediate fields, far below the arrival of PE region. The crossover in Ref. 22 is designated as the one from weak to strong pinning. However, we believe that in our single crystal of YNi$_2$B$_2$C the hump at $h \sim 0.2$ is the mark of the pining crossover from weak to still weaker pining.

### C. Anisotropy in vortex phase diagram

Collating all the data, we can now construct the vortex phase diagram for YNi$_2$B$_2$C crystal in the basal plane. Figures 6 (a) and 6(b), respectively, show the phase diagrams for YNi$_2$B$_2$C along [100] and [110] directions. The observation of SMP in [100] direction gives an additional line in its phase diagram. It seems that the ordered Bragg glass extends over a large part of the ($H,T$) phase space up to the onset of the PE for $H \parallel [110]$ and, thereafter it crosses over to the completely disordered region ($> H_p^{on}$). However, for $H \parallel [100]$, one can see that the amorphous state is reached via two steps; in the first step, a large domain defining the ordered Bragg-glass state transforms to the multi-domain vortex glass state across $H_{SMP}^{on}$. In the second step, the multi-domain vortex glass state crosses over to the amorphous region via $H_p^{on}$ line in the phase diagram. Amorphous region is often [23] considered to subdivide in to two parts separated by the $H_p$ line between the $H_p^{on}$ and $H_{c2}$ line. The disordering process which starts at the onset field of the PE, i.e., at $H_p^{on}$, probably gets completed at the peak field $H_p$, producing a completely amorphous state. Hence, the vortex phase diagram can be considered to comprise [23, 24] a Bragg glass ($H_{SMP} < H < H_{SMP}^{on}$), a multi-domain (elastic) vortex glass ($H_{SMP}^{on} < H < H_p^{on}$), a plastic glass ($H_p^{on} < H < H_p$), and a completely amorphous region ($H_p < H < H_{c2}$). We also show variation of $H_{c2}$, $H_p$ and $H_p^{on}$ as a function of the angle at 2 K in the inset of Fig. 6(b), as the field is rotated in the basal plane of the crystal. One can observe considerable amount of change in the values of $H_{c2}$, $H_p$ and $H_p^{on}$ as a function of the angle. The angular variation in the phase diagram suggests that its details depend critically on the field-angle in which the FLL is prepared.

## 4. CONCLUSION

To summarize, we have presented the experimental data pertaining to the angular dependence of the dc magnetization hysteresis for a weakly pinned YNi$_2$B$_2$C single crystal, by rotating it within the basal plane ($H \parallel [100]$ to $H \parallel [010]$) and out-of the basal plane (from $H \parallel c$ to $H \perp c$) with respect to the direction of the magnetic field. The PE phenomenon is observed for all crystallographic orientations, but, the SMP anomaly is observed only for $H \parallel a$. The PE inferred from the normalized critical current density ($j_c$) suggests that the ordered FLL for $H \parallel [110]$ spans a larger field interval within the basal plane and, therefore the FLL is better ordered. However, best ordered FLL is found for $H \parallel c$. It seems that perfect square FLL symmetry attainted for $H \parallel c$ promotes the realization of better spatial order prior to the onset of the PE. When the quasi-square FLL is formed within the basal plane, the better ordered FLL is found along the $H \parallel [110]$. We also found a weak temperature dependence of the pinning force density and the shape of the pinning force density ($F_p$) changes along the different crystallographic orientations. Other than the PE and a usual pinning force maximum, a third maximum is observed in the intermediate field range ($h \sim 0.2$), which is also associated with the sharper drop in $j_c$, probably marking the crossover from a weak to still weaker pinning regime.

# References

1. H. Küpfer, G. Linker, G. Ravikumar, T. Wolf, A.A. Zhukov, R. Meier-Hirmer, B. Obst, and H. Wühl, *Phys. Rev. B* **67** (2003), p. 064507.
2. M. Yethiraj, D.K. Christen, D.McK. Paul, P. Miranovic, and J.R. Thompson, *Phys. Rev. Lett.* **82** (1999), p. 5112.
3. T. Park, E.E.M. Chia, M.B. Salamon, E.D. Bauer, I. Vekhter, J.D. Thompson, E.M. Choi, H.J. Kim, S.I. Lee, and P.C. Canfield, *Phys. Rev. Lett.* **92** (2004), p. 237002.
4. D. Jaiswal-Nagar, T. Isshiki, N. Kimura, H. Aoki, S. Ramakrishnan, and A.K. Grover, *Physica C* **460-462** (2007), p. 1215.
5. D. Jaiswal-Nagar, Vortex State Studies in Borocarbide Superconductors: $YNi_2B_2C$ and $LuNi_2B_2C$, PhD thesis, Tata Institute of Fundamental Research, Mumbai (2006).
6. P. Raychaudhuri, D. Jaiswal-Nagar, G. Sheet, S. Ramakrishnan, and H. Takeya, *Phys. Rev. Lett.* **93** (2004), p. 156802.
7. C.P. Bean, *Rev. Mod. Phys.* **38** (1964), p. 36.
8. W.A. Fitez and W. W. Webb, *Phys. Rev.* **178** (1969), p. 657.
9. G. Ravikumar, P.K. Mishra, V.C. Sahni, S.S. Banerjee, A.K. Grover, S. Ramakrishnan, P.L. Gammel, D.J. Bishop, E. Bucher, M.J. Higgins, and S. Bhattacharya, *Phys. Rev. B* **61** (2000), p. 12490.
10. S.S. James, C.D. Dewhurst, R.A. Doyle, D.McK. Paul, Y. Paltiel, E. Zeldov, A.M. Campbell, *Physica C* **332** (2000), p. 173.
11. X.S. Ling, S.R. Park, B.A. McClain, S.M. Choi, D.C. Dender, and J.W. Lynn, *Phys. Rev. Lett.* **86** (2001), p. 712.
12. T. Nishizaki and N. Kobayashi, Supercond. Sci. and Technol. **13** (2000), p. 1.
13. A.I. Larkin and Y.N. Ovchinnikov, *Sov. Phys. JETP* **38** (1974), p. 854.
14. A.I. Larkin and Y.N. Ovchinnikov, *J. Low Temp. Phys.* **34** (1979), p. 409.
15. S.S. Banerjee, S. Ramakrishnan, A.K. Grover, G. Ravikumar, P.K. Mishra, V.C. Sahni, C.V. Tomy, G. Balakrishnan, D.Mck. Paul, P.L. Gammel, D.J. Bishop, E. Bucher, M.J. Higgins, and S. Bhattacharya, *Phys. Rev. B* **62** (2000), p. 11838.

16. M.R. Eskildsen, P.L. Gammel, B.P. Barber, A.P. Ramirez, D.J. Bishop, N.H. Andersen, K. Mortensen, C. A. Bolle, C. M. Lieber, and P. C. Canfield, *Phys. Rev. Lett*. **79** (1997), p. 487.
17. M.R. Eskildsen A.B Abrahamsen, D. López, P.L. Gammel, D.J. Bishop, N.H. Andersen, K. Mortensen, and P. C. Canfield, *Phys. Rev. Lett*. **86** (2001), p. 320.
18. H. Sakata, M. Oosawa, K. Matsuba, N. Nishida, H. Takeya, and K. Hirata, *Phys. Rev. Lett*. **84** (2000), p. 1583.
19. V. Metlushko, U. Welp, A. Koshelev, I. Aranson, G.W. Crabtree, P.C. Cafield *Phys. Rev. Lett*. **79** (1997), p. 1738.
20. A.V. Silhanek, J.R. Thompson, L. Civale, D.Mck. Paul, and C.V. Tomy, *Phys. Rev. B* **64** (2001)**,** p. 012512.
21. D. Jaiswal-Nagar, A.D. Thakur, D. Pal, S. Ramakrishnan, A.K. Grover, and H. Takeya, *Phys. Rev. B* **74** (2006), p. 184514.
22. S. Mohan, J. Sinha, S.S. Banerjee, and Y. Myasoedov, *Phys. Rev. Lett*. **98** (2007), p. 027003.
23. S.S. Banerjee, N.G. Patil, S. Ramakrishnan, A.K. Grover, S. Bhattacharya, P.K. Mishra, G. Ravikumar, T.V. Chandrasekhar Rao, V.C. Sahni, C.V. Tomy, G. Balakrishnan, D.Mck. Paul, *Rev. B* **59** (1999), p. 6043.
24. A.D. Thakur, S.S. Banerjee, M.J. Higgins, S. Ramakrishnan, and A.K. Grover, *Phys. Rev. B* **72** (2007), p. 134524.

# FIGURE CAPTIONS.

FIGURE 1.
Isothermal magnetization hysteresis loop (the first and second quadrant) at 2 K for $H \parallel [100]$. Inset panels (a) and (b), respectively, show the expanded portions to highlight the peak effect (PE) and second magnetization peak (SMP).

FIGURE 2.
Part of the magnetization hysteresis loop at (a) 2 K and (b) 6 K for different orientations of the crystal with respect to the magnetic field direction. The insets show the expanded portions near the PE region.

FIGURE 3.
Variation of critical current density $j_c$ as a function of reduced field $h$ ($= H/H_{c2}$) on a log-log scale for the different orientations of the crystal in the basal plane with respect to the filed. In panel (a), the current density is normalized with respect to its value at the peak of the peak effect whereas the normalization is with respect to the $j_c$ value at low fields (200 Oe) in panel (b).

FIGURE 4.
Variation of critical current density $j_c$ as a function of reduced field $h$ ($= H/H_{c2}$) on a log-log scale for rotations of the crystal out of the basal plane with respect to the filed. $j_c$ is normalized with respect to its value at the peak of the peak effect. The dotted line near h ~ 0.2 appears to identify the start of the crossover region separating the two power law regimes in $j_c$ vs. h for $H \parallel c$.

FIGURE 5.
Normalized pinning force density as a function of reduced field $h$ for $H \parallel [100]$ {panel (a)} and $H \parallel [110]$ {panel (b)} at different temperatures. In panel (c) we show the same for different orientations of the crystal with respect to the magnetic field at 2 K.

FIGURE 6.
The ($H,T$) phase diagram for a YNi$_2$B$_2$C crystal in the basal plane, (a) along [100] and (b) along [110]. Different vortex states are marked. The inset in panel (b) shows the variation of the onset of peak position ($H_p^{on}$), peak of the peak position ($H_p$) and upper critical field ($H_{c2}$) as a function of rotation angle, i.e., the angle between the magnetic field and the crystallographic direction of the crystal.

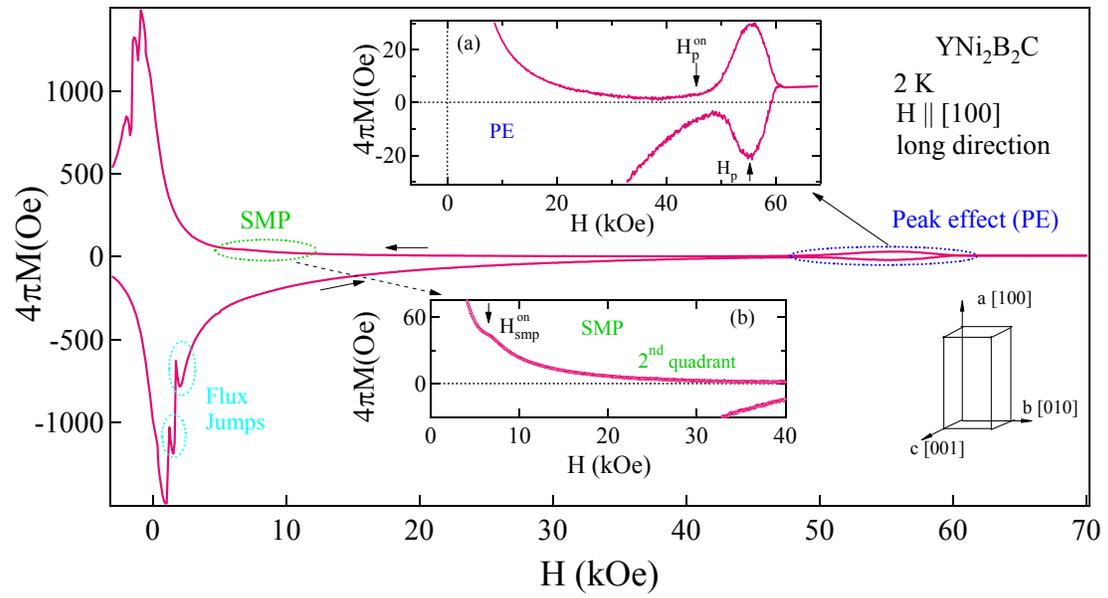

**Fig. 1**

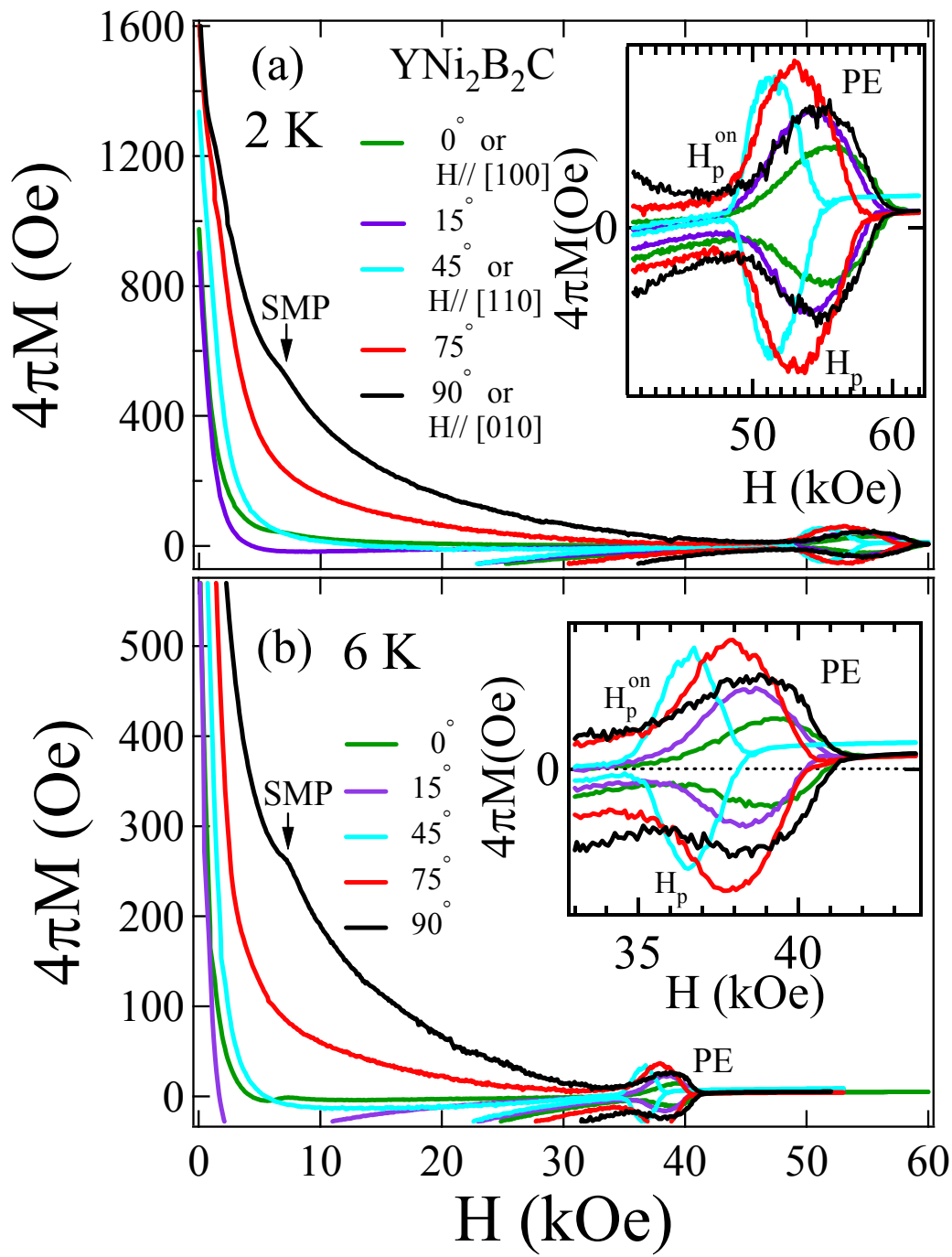

**Fig. 2**

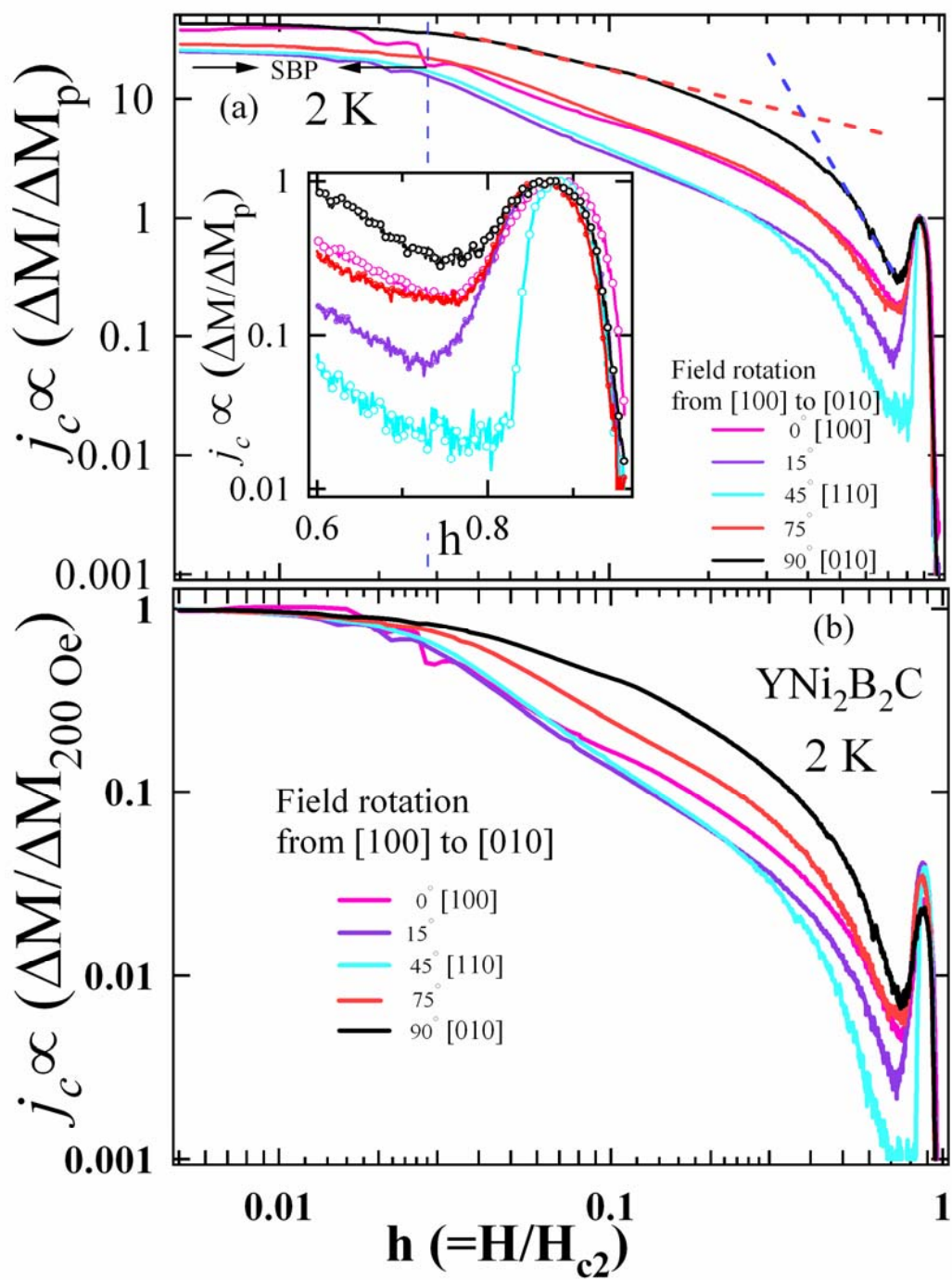

**Fig. 3**

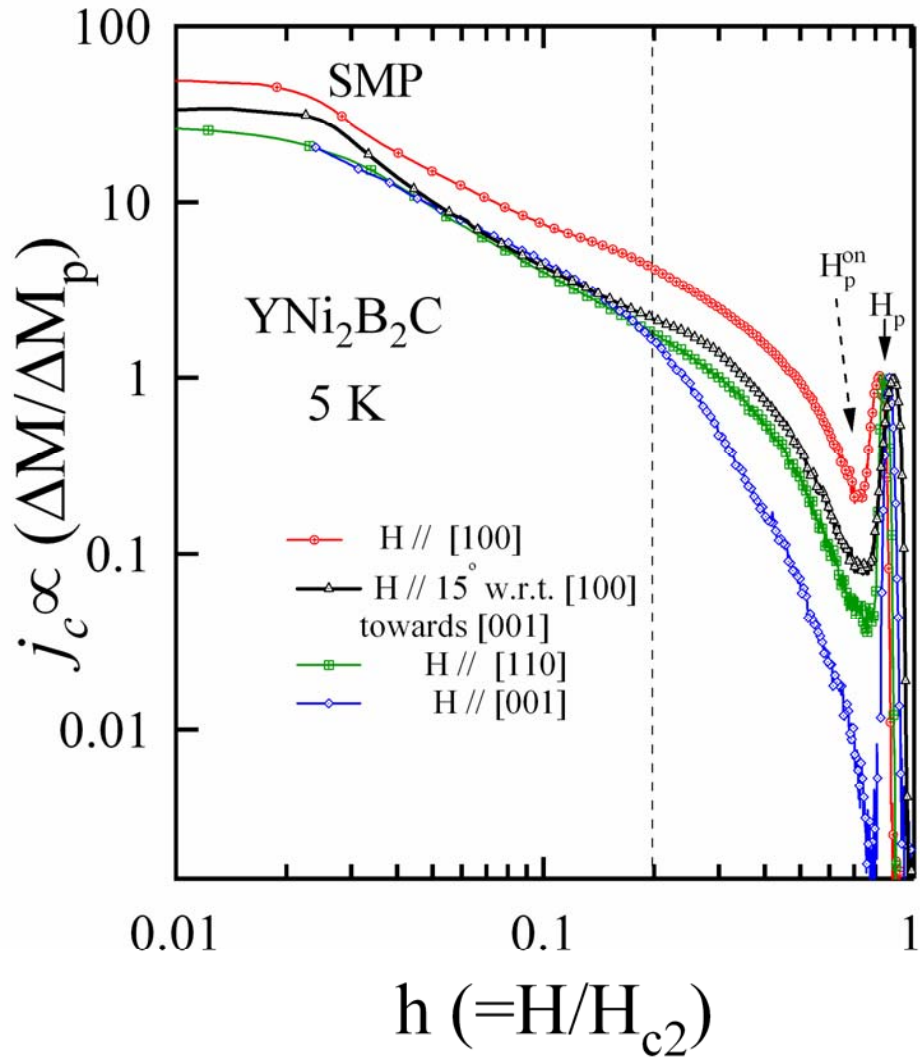

**Fig. 4**

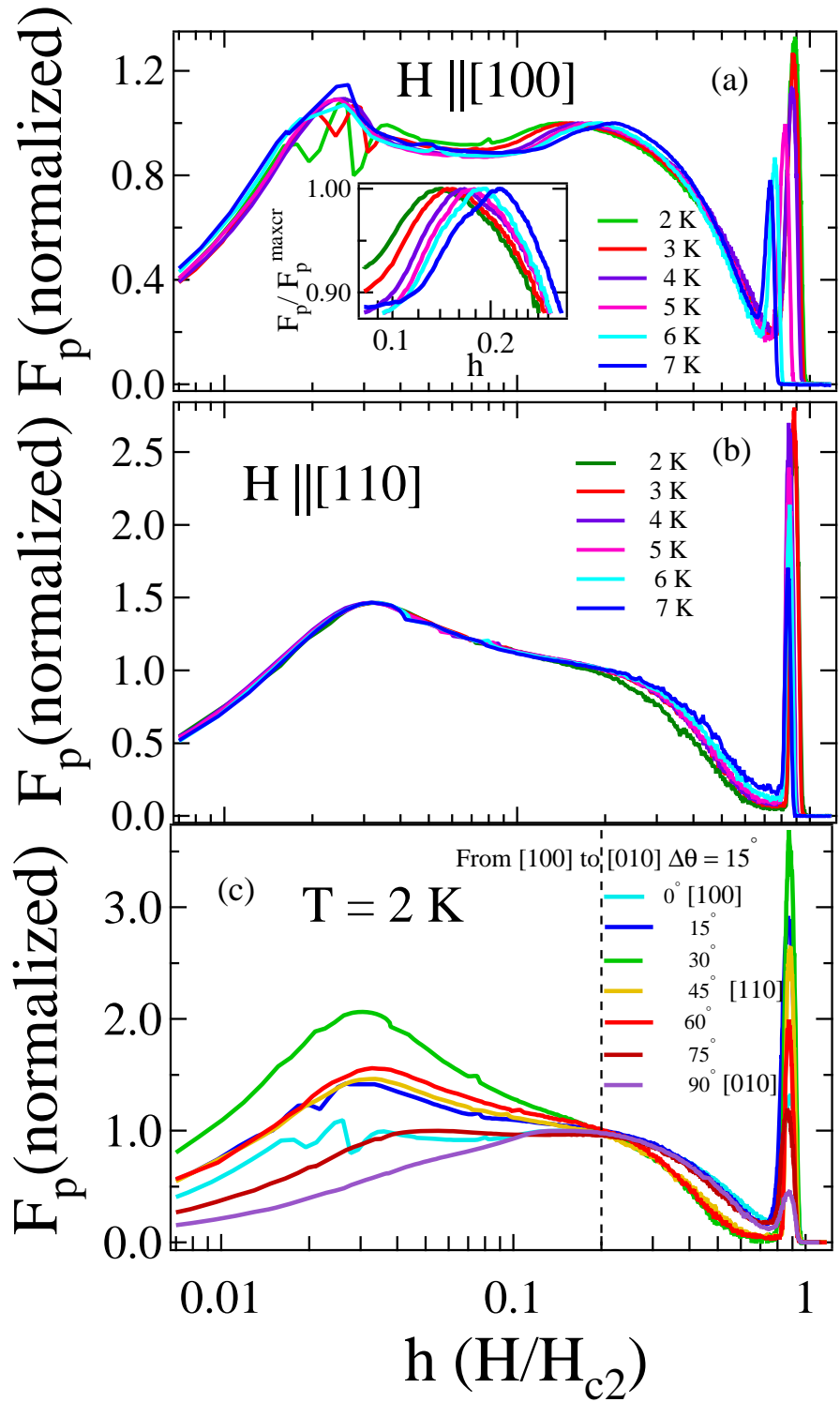

**Fig. 5**

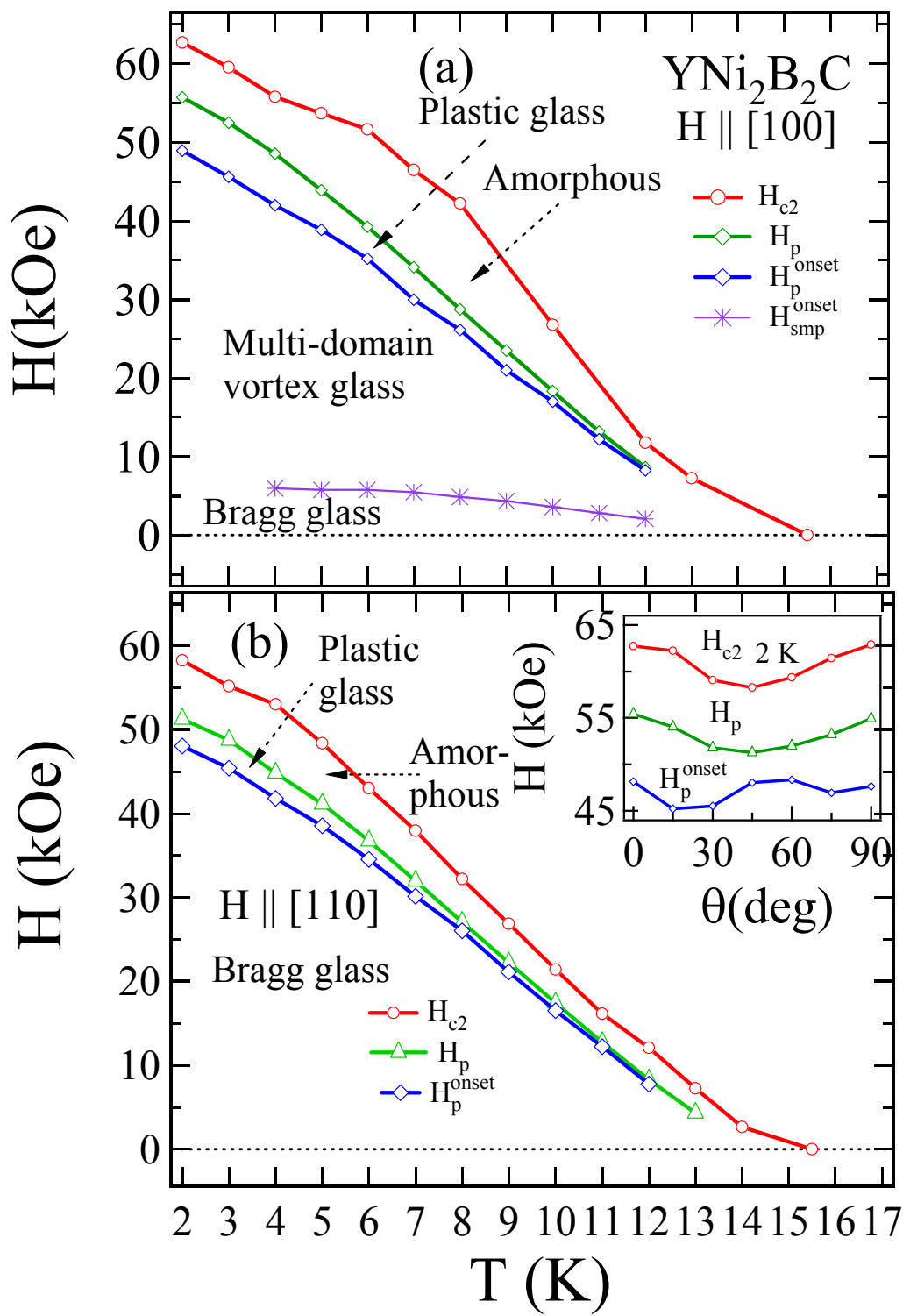

Fig. 6